\documentstyle[12pt,psfig]{article}
\begin{document}

\vskip 1truecm
\rightline{Preprint  PUPT-1681, DAMTP HEP/97-17}
\rightline{ e-Print Archive: hep-ph/9703266}
\vspace{0.2in}
\centerline{\Large Lattice Chern-Simons Number Without 
	Ultraviolet Problems} 
\vspace{0.3in}
\centerline{\Large Guy D. Moore\footnote{e-mail:
guymoore@puhep1.princeton.edu }
}
\medskip

\centerline{\it Princeton University}
\centerline{\it Joseph Henry Laboratories, PO Box 708}
\centerline{\it Princeton, NJ 08544, USA}

\vspace{0.2in}

\centerline{\Large Neil Turok \footnote{e-mail: 
N.G.Turok@damtp.cam.ac.uk }
}
\medskip

\centerline{\it DAMTP, Silver Street}
\centerline{\it Cambridge, CB3 9EW, UK}

\vspace{0.2in}

\centerline{\bf Abstract}

We develop a topological method of measuring Chern-Simons number change
in the real time evolution of classical lattice SU(2) and SU(2) Higgs
theory.  We find that the Chern-Simons number diffusion rate per
physical 4-volume is very heavily suppressed in the broken phase, 
and that it decreases with lattice spacing in pure 
Yang-Mills theory, although not as quickly
as predicted by Arnold, Son, and Yaffe.




\section{Introduction}
\label{introduction}

In pioneering work Andrei
Sakharov \cite{Sakharov} pointed out that
unified theories of elementary particle physics had the potential to
explain one of cosmology's basic mysteries, namely
the  overabundance of matter over antimatter in the universe. 
Since then, two main embodiments of this idea have emerged. 
Early work on the subject attempted to exploit the 
baryon number violation present in grand unified theories (GUTs), and 
this is still a popular approach. The idea here was simple, 
that particles whose decays violated baryon number might fall out
of equilibrium as the universe cooled, and their decays could lead
to a baryon asymmetry.  However, as usual with GUTs the main problem is
an overabundance of models 
and associated free parameters, which are difficult to 
experimentally constrain since most of the parameters are 
relevant to physics at inaccessibly high energies. 
Furthermore, the failure of experiments to 
detect evidence for GUT baryon number violation makes 
the prospect of experimental test seem at present remote.

The last decade has seen much more interest in the idea that much lower 
energy physics, within reach of particle accelerators and other experiments,
might yet realize Sakharov's vision. It has
been known since the work of t'Hooft that baryon number is violated in
the Minimal Standard Model \cite{tHooft}, for deep reasons relating to
a) its chirality and b) the topology of the vacuum. The 
baryon number current $J^B_\mu$ is not conserved because only the 
left handed fields couple to the SU(2) gauge fields, and their 
fermion number current possesses an axial anomaly: 
\begin{equation}
\partial^\mu J^B_\mu = \frac{- N_F g^2}{32 \pi^2} {\rm Tr} \tilde{F}^{\mu \nu}
	F_{\mu \nu} = \frac{ - N_F g^2}{32 \pi^2} \partial^\mu K_\mu \, .
\end{equation}
The charge associated with this anomaly, 
$(g^2/32\pi^2) \int d^3 x K_0 \equiv
N_{CS}$, the Chern-Simons number, can take on any integer value in the 
vacuum because the gauge group SU(2) has a nontrivial third homotopy group.
(Note that we have ignored the hypercharge term in equation (1), because its
Chern-Simons number is forced to be zero in the vacuum, so it can never
contribute to a `permanent' change in the baryon number).

At first the violation of baryon number in the standard model
was considered an irrelevant peculiarity, because at low energies,
changes to $N_{CS}$ only occur due to quantum mechanical 
tunneling, which is exponentially suppressed by
$O(\exp(-4 \pi/\alpha_W)) \sim \exp(-400)$.
However, the rate need not be so suppressed at higher temperatures
\cite{KRS85}.  More careful consideration shows that transitions 
between vacua of differing $N_{CS}$ can be excited thermally at a rate
which in the broken electroweak 
phase is dominated by trajectories which pass near
to the minimal energy ``Sphaleron'' saddlepoint \cite{Manton}, and the 
exponential suppression is $\sim \exp(-E_{sph}/T) =
\exp(-2 B M_W(T)/\alpha_W T)$, with $M_W(T) = g_W v(T)/2$, 
$v(T)$ the size of the Higgs condensate at temperature $T$, and
$B$ a constant ranging from 1.5 to 2.7  as the zero temperature Higgs mass
ranges from zero to infinity \cite{ArnoldMcLerran}.

However, this semiclassical calculation breaks down in the
symmetric electroweak phase.  Here the Higgs condensate has dissolved and 
there is no barrier to thermally activated
$N_{CS}$ change, and hence to baryon number violation. 
Infrared gauge field configurations are expected to dominate
the diffusive behavior of $\int d^4 x \tilde{F}^{\mu \nu} F_{\mu \nu}$, 
because the energy of intermediate 
field configurations which the system must pass through to permanently
change $N_{CS}$, 
carrying half-integral $N_{CS}$, scales inversely with their size. 
Thus short wavelength (UV) contributions to $\tilde{F}^{\mu
\nu} F_{\mu \nu}$ are expected to behave oscillatorily and not permanently
change $N_{CS}$.  In this case the diffusion of $N_{CS}$ is believed
to be essentially classical since it involves high occupation number
infrared fields, which should conform closely to the classical field
approximation. In the classical field theory at finite temperature there
are two scales, namely the natural nonperturbative
length scale  $1/(\alpha_W T)$ and the (lattice) regulator or cutoff scale. 
If one assumes that the long wavelength
phenomena are independent of the cutoff scale (which is still unclear)
then dimensional analysis indicates the rate of Chern Simons number diffusion
behaves parametrically as
\begin{equation}
\Gamma \equiv \lim_{t \rightarrow \infty} \frac{ \langle (N_{CS}(t)
	- N_{CS}(0) )^2 \rangle }{V t} = \kappa (\alpha_W T)^{4}
\end{equation}
with $\kappa$ an unknown constant of order unity.

There are no reliable analytical estimates of $\kappa$, 
and at this time it appears that the
only hopeful way of determining $\kappa$ is by numerical, lattice study
of the classical theory, which should be a suitable analog theory in
the infrared 
\cite{GrigRub,Ambjornearly,BodMc,AmbKras,TangSmit,MooreTurok,Bodeker}.

There has already been substantial work on finding $\kappa$ numerically
\cite{Ambjornearly,Ambjornetal,AmbKras,Moore1,Moore2,TangSmit,MooreTurok}.
All this work has been based on a lattice implementation of $N_{CS}$ 
as an integral over a local operator, usually
\begin{equation}
2 \pi^2 \frac{dN_{CS}}{dt} = \sum_{ {\rm sites} \: x} 
	\sum_{ {\rm directions} \: i} \left[ \left( 
	\frac{ E^a_i(x) + E^a_i(x-\hat{i}) }{2} \right) \times 
	\frac{1}{4} \sum_{\Box_{jk}} \left( \frac{1}{2} {\rm Tr} -i \tau^a
	U_{\Box_{jk}} \right) \right]
\label{olddef}
\end{equation}
(appropriate parallel transports implied)
where the $E^a_i$ are the electric fields entering and leaving the point
in the $i$ direction, and the latter sum is over the four plaquettes 
orthogonal to the $i$ direction and with a corner at the point $x$.  That
is, one takes a sum over lattice sites of $E \cdot B$, with $E$ and $B$
the average over neighboring space-time and space-space plaquettes of
the Lie algebra content of the plaquette.  This definition is chosen
to reproduce $(g^2/32 \pi^2) \int d^4 x {\rm Tr}\tilde{F}^{\mu \nu} 
F_{\mu \nu}$ for smooth, slowly varying gauge fields, which it does.
However, as a definition of the time derivative of 
$N_{CS}$ it has several drawbacks.  First, it is not a total 
time derivative \cite{Moore1}.  Second,
it possesses white noise which grows as $1/a$ (with $a$ the lattice
spacing) \cite{AmbKras}.  It also gives a UV dominated, `fake' contribution
to the inferred diffusion of $N_{CS}$, which is clearly seen
in the broken phase  \cite{MooreTurok,ArnoldYaffe,AmbKras2}. Finally, UV- 
dominated (tadpole) 
contributions from higher order terms in the expansion of $U$ may 
renormalize the inferred rate \cite{TangSmit}.  These drawbacks are 
serious, and until they are resolved we cannot have much confidence in the
derived values of $\kappa$, particularly in the broken phase where the
true rate will be smaller and any spurious UV generated diffusion
could dominate.

We believe it is 
important to clear up these problems in order to get a valid 
determination of the efficiency of baryon number violation, which is
simply related to $\kappa$ \cite{KhlebShap,RubShap}.  In particular, it
is not known whether $\kappa$ will depend on the physics of charge
screening (hard thermal loops).  The quantum theory has in addition to
its natural nonperturbative length scale $l_n \propto 1/(g^2 T)$ a
Debye screening length $l_d \propto 1/(gT)$, and it
has recently been argued that $\kappa$ should depend on the ratio of $l_d$
and $l_n$ as $\kappa \propto (l_d/l_n)^2$ \cite{ArnoldYaffe,HuetSon}. 
Classical lattice simulations mimic 
hard thermal loop effects through interactions between the infrared
modes and the short wavelength lattice modes
\cite{BodMc}, giving an effective screening 
length proportional to the square root of the lattice spacing. 
If hard thermal loops contribute as claimed, 
one should see  $\kappa$ decrease linearly with lattice spacing.
At present the numerical data argues against such
a dependence \cite{AmbKras,Moore1}, but the evidence 
cannot be considered strong until the problems in the definition of 
$N_{CS}$ have been dealt with.  
Similarly, the problem of hard thermal
loops in general \cite{BodMc} may be addressable by adding particle
species which generate hard thermal loop effects to the classical
simulations \cite{chaoran}, but pursuing this project also demands a good 
definition of $N_{CS}$.

In this paper we present an alternative, topological definition 
of $N_{CS}$ on the 
lattice, related to that of Woit \cite{Woit}.  Its change as the system
evolves smoothly from one configuration to another 
is strictly independent of
small variations of the path between configurations, 
which prevents
UV generated diffusion.  It is also gauge invariant under small 
gauge transformations. The technique is to add a group-valued 
scalar field in the fundamental representation, which lives in the
background of the gauge fields but does not influence their evolution.
The role of this field is to track topology change in the gauge field, and 
$N_{CS}$ is to be associated with minus the winding number of the minimal
energy configuration of this ``slave'' field.  We outline the general idea in
Section \ref{two}, and address the particulars of its lattice implementation
in Section \ref{three}.  In Section \ref{four} we present numerical results
for this technique in Yang-Mills and Yang-Mills Higgs theory.  Section
\ref{conclusion} concludes.  

\section{Slave field in the continuum}
\label{two}

Consider a continuum SU(2) spatial gauge field configuration $A_i^a(x)$.
We want a surrogate for its Chern-Simons number $N_{CS}$ which will be
easy to implement on the lattice.  Our choice is to add a notional
SU(2) valued
scalar field $S(x)$.
Under a gauge transformation $S(x)$ tranforms as
$S(x) \rightarrow g(x)S(x)$.  $S$ will be a ``slave'' field, meaning that
it lives in the background of the gauge connection but it will not enter
the dynamics of the gauge fields or any other physical fields in the
problem.  It may however be used to change the gauge in which we are working,
in order to find the gauge in which $A_i^a(x)$ is smoothest.

Any SU(2) matrix may be expressed as
$S_{11}= a$, $S_{12}= b$, 
$S_{22}= a^*$, $S_{21}= -b^*$, with $a=x^0 + i x^3$
and $b= x^2 + i x^1$, and $x^i$ a real four-component vector of
unit length, $x^i x^i =1$. Thus the
the slave field $S(x) $ provides a map from space to SU(2)=$S^3$.
The winding number of this map is
\begin{equation}
N_S = {1\over 24 \pi^2} \int d^3 x \epsilon^{ijk} {\rm Tr} \left(
\partial_i S S^{\dagger} \partial_j S S^{\dagger} \partial_k S S^{\dagger}
\right)
\end{equation}
which as we shall see may be very efficiently computed on the lattice.
We shall give the slave field the Hamiltonian
\begin{equation}
H(S) = \int d^3 x {\rm Tr} (D_i S)^{\dagger} (D_i S)
\end{equation}
and note the following property:  if the gauge connection is a vacuum 
(pure gauge) connection
then the minimum energy configuration for the $S$ field
has a winding number $N_S$ equal to minus the Chern-Simons number of the gauge 
field configuration.
To see this, note
first that in the gauge where $A_i^a=0$ everywhere, the minimum energy 
configuration for $S(x)$ is $S(x) = I$ everywhere (or any fixed
element of SU(2)).  This configuration has zero winding number and zero
energy.  By gauge invariance, the minimum energy configuration for $S$ in
any other vacuum configuration $A_i^a(x)$ must also have zero energy.
If we gauge transform by $g(x) = S^{\dagger}(x)$, which carries $S(x)$ 
to $I$, the Hamiltonian then measures $\int A^2$, so the gauge field 
in this gauge must be $A_i^a(x) = 0$ everywhere.  Thus $S^{\dagger}(x)$
is the gauge transformation which carries $A_i(x)$ into the trivial vacuum.
The Chern-Simons number of a vacuum gauge field is just the 
winding number of the gauge change which carries it to $A_i=0$, which in this
case is the winding number of $S^{\dagger}(x)$; and
$S(x)$ has minus this winding number, 
since $N_S$ is reversed by complex conjugation.
Thus the winding number of the minimal energy configuration for $S(x)$ is
$-N_{CS}(A)$.

On any compact space with smooth connection $A_i^a(x)$, it should always be 
possible to find a minimal energy slave field configuration (modulo
a global SU(2) rotation), and we can take minus its winding number,
$-N_S$, as our surrogate for Chern-Simons number\footnote{There could be
multiple absolute minima but the definition would only be ambiguous in 
the unlikely event that two had different winding number.}.  By no means does
$-N_S$ always equal $N_{CS}(A)$; in particular $N_S$ is always an integer,
unlike $N_{CS}$; but it shares these important properties with $N_{CS}$:
\begin{itemize}
\item it is a function only of the configuration and not of the path to
	the configuration,
\item it is invariant under small gauge changes and changes by an integer,
	equal to the winding number of the gauge group element $g(x)$, 
	under large ones,
\item it coincides with $N_{CS}$ for vacuum configurations.
\end{itemize}
Since we are only interested in the long time diffusion 
of $N_{CS}$ across the set of gauge equivalent configurations,
we only care about the topological information specified by $N_{CS}$.
And for this, 
$-N_S$ carries all of the appropriate information.
One can make this argument
rigorous by considering the case
where one begins in vacuum, heats up  the 
gauge fields and evolves them for a very long time before 
cooling down to vacuum
again. In this situation the change in $-N_S$ would be 
precisely equal to the
change in $N_{CS}$, and clearly in the long time limit 
the inferred diffusion rate 
would be independent of the initial heating and cooling times.

In practice the gauge fields $A_i(x)$ will evolve, and it is too much
work to continually find the minimum energy solution for $S(x)$.  It is
more practical to give $S$ very efficient dissipative dynamics so that
it will always remain close to an energy minimum.  How will these 
dissipative dynamics behave when the gauge field configuration changes
from one winding number vacuum to another?  To get a general idea it
is useful to consider the dissipative dynamics of $S(x)$ when it is in
the wrong topological sector for the underlying gauge fields.  For instance,
suppose that $S(x) = I$ everywhere but the gauge field is in
a winding 1 vacuum configuration.
To see how $S(x)$ evolves, it is convenient to change gauge so that
$A_i^a(x)=0$ but $S(x)$ is in a winding -1 configuration.  This is just
a standard nonlinear sigma model on $S^3$ in 3 dimensions, initially 
in a winding -1 configuration, and its evolution is 
is well known \cite{probablyNeil}.
The winding configuration (or texture) is unstable to collapse; its potential 
energy decreases linearly with its size. 
So the winding in the slave field shrinks and 
concentrates until a singularity (a texture unwinding event) occurs. 
The winding then jumps discontinuously to the correct
value, allowing the system to settle down to the correct minimum energy
configuration.  The important point to make is that there is no energy
gap between topological sectors for the slave field; by making a nontrivial
winding of arbitrarily small spatial extent one can get nonzero winding at 
arbitrarily little energy cost.  Hence topology does not present an
obstruction to the dissipative evolution finding the appropriate winding
number, minimal energy configuration.  Secondly, the winding number change
occurs by the concentration of winding in a very small
region and the development of a singularity.

What the slave field technique does is find the gauge 
transformation $S^{\dagger}(x)$ which minimizes $\int \vec{A}_S^2$
with respect to $S(x)$, where $\vec{A}_S$ is the gauge transform 
of  $\vec{A}$ by the gauge group element $S^{\dagger}(x)$ 
($S(x)$ the slave field).  We then take
$N_{CS}$ to be the winding number of that gauge transformation. 
In doing so, we make
the approximation that the Chern-Simons number of the configuration with
minimal $\int \vec{A}^2$ is zero.  Obviously this is not correct, but the
value of $N_{CS}$ for this configuration should be modest and
in particular will not grow without limit over the course of a Hamiltonian
evolution of the gauge fields.  Therefore, while the slave field
definition will have some
noise about the correct value of $N_{CS}$, due to the approximation just
mentioned, the noise will have no diffusive
power so the diffusion constant of $N_{CS}$ will
be reproduced correctly by this procedure.  

Another concern is that, by finding the slave field configuration through
dissipative dynamics, we are not guaranteed to find the global minimum of
$\int \vec{A}^2$ in the space of gauges, but only a local minimum.
For nontrivial
gauge fields there may be a Gribov ambiguity and multiple local
extrema \cite{Gribov}.  In our case this is only a problem if at least one
of the extra extrema is a minimum (within the Gribov horizon) with a 
different winding number than the absolute minimum.  This is possible.
However, we do not expect that, in finite volume
and with gauge fields which are smooth at short distances,
such minima should exist with a large winding number 
difference from the global minimum.  In this case the possibility of
finding such a false minimum will increase the ``noise'' in this measurement
of $N_{CS}$ (the typical difference between the measured winding number and
the true Chern-Simons number now being the Chern-Simons number of the 
typical Gribov copy), but it will not lead to excess diffusive power during a 
long evolution of the gauge fields (since the size of the error will
be bounded and will not grow with time).  We will see this explicitly
in the next section, when we compare this winding number measurement
technique with an alternate ``improved'' technique due to Ambjorn and 
Krasnitz \cite{AmbKras2}.

Finally, we mention another interesting consequence of the slave field
transformation, which is that when the minimum energy configuration is
found, $S(x)$ 
 obeys $D_i D_i S=0$.  Upon gauge transforming to $S(x)=I$, this becomes
the Coulomb gauge condition $\partial_i A_i =0$. So the slave field
technique is in fact just a way of determining the gauge change to
Coulomb gauge.

\section{Implementing the slave field on the lattice}
\label{three}

We now turn to the question of implementing this idea on the lattice.

\subsection{defining the slave field and winding number}

It is clear how to define the slave field $S$ on the lattice.  It will take
on a value $S(x) \in $SU(2) at each lattice point which will transform 
under a gauge change as $S(x) \rightarrow g(x)S(x)$, and the parallel
transport of $S(x)$ to the point $x-i$ will be $U_i(x-i)S(x)$.  The 
simplest lattice Hamiltonian which gives the right continuum limit is
\begin{equation}
H = \sum_{x,i} \left( 1 - \frac{1}{2} {\rm Tr} S^{\dagger}(x) 
	U_i(x) S(x+i) \right) 
  = \sum_{x,i} \left( 1 - \frac{1}{2} {\rm Tr} S^{\dagger}(x+i) 
	U_i^{\dagger}(x) S(x)  \right)\, ,
\end{equation}
and the winding number can be defined by a construction similar to that
of Woit \cite{Woit}, as follows.

First we partition the lattice into unit cubes with lattice points as
vertices.  Each cube has a basepoint, and extends one unit in the $+x$, 
$+y$, and $+z$ directions from the basepoint; so the vertices of the 
cube are $(i,j,k)$, $(i+1,j,k)$, $\ldots$ $(i+1,j+1,k+1)$ with $(i,j,k)$
the basepoint. A cube is called even if the sum of the coordinates of the 
basepoint is even and odd otherwise.  Next we partition each cube into 5
tetrahedra, as shown in Figure \ref{triangulate}.  The triangulation is
such that the faces of the tetrahedra match, i.e. we have represented 
three dimensional space
as a simplicial complex.  Consider the slave field $S(x)$ to take its
values in the 3-sphere, using the identification of the last section.
In each tetrahedron, we will interpolate the
function $S(x)$ between its values on the vertices by a geodesic rule.  
For each pair of vertices on a tetrahedron we draw on the 3-sphere a
great circle connecting the image points, 
and take $S(x)$ on the line connecting the two
vertices to take on the values along the smaller arc of this great circle.
Each face of a tetrahedron now has its edges map into a geodesic triangle
on the 3-sphere; take the face of the tetrahedron to fill in the inside
of the triangle (the geodesic triangle breaks the equatorial sphere defined
by the 3 points into two regions, and we fill in the region of smaller
area).  Finally, the surface of each tetrahedron now encloses a geodesic
tetrahedron on the 3-sphere; we take the interior of the tetrahedron to
fill in the inside of this tetrahedron on the 3-sphere (the surface of the
tetrahedron maps into a surface which breaks the 3-sphere into two 
regions, and we take the interior of the tetrahedron to fill 
in the smaller volume region).  This construction is unique except for
a set of measure zero (when the convex hull of the verticies of some
tetrahedron contains the origin) and it is easy to determine 
the winding number of this interpolated $S(x)$.

The winding number of a map from some space (in this case the 3-torus on
which the lattice lives) into the 3-sphere can be evaluated (for 
piecewise smooth
maps which have nonzero Jacobian off a set of measure zero) by the following
simple algorithm \cite{Woit}; choose a point $q$ on the 3-sphere, 
and find its inverse images.  The winding 
number is the sum over inverse images of $q$ of the sign of
the Jacobian at that point, i.e. the number of times $q$ is covered with
positive orientation minus the number of times it is covered with negative
orientation.  In our case, this means that we must choose a point 
$q$ on the 3-sphere, and for each tetrahedron we must determine whether
the tetrahedron covers that point on the 3-sphere, and if so with what
orientation.  If the four vertices $v_1,v_2,v_3,v_4$ of the tetrahedron 
map into the points $p_1,p_2,p_3,p_4 \in S^3$ (note that $v_i$ 
must be positively oriented, i.e. $( v_4- v_1) \cdot 
( (v_1 - v_2) \times (v_2 - v_3) ) > 0$), then the orientation of the 
image of the tetrahedron is
\begin{equation}
{\rm sign}( \epsilon_{ijkl} p_1^i p_2^j p_3^k p_4^l ) = {\rm orientation}
\end{equation}
and the point $q\in S^3$ lies inside the image of the tetrahedron if
and only if
\begin{eqnarray}
{\rm sign}( \epsilon_{ijkl} q^{i}   p_{2}^j p_{3}^k p_{4}^l ) 
	& = & \nonumber \\
{\rm sign}( \epsilon_{ijkl} p_{1}^i q^{j}   p_{3}^k p_{4}^l ) 
	& = & \nonumber \\
{\rm sign}( \epsilon_{ijkl} p_{1}^i p_{2}^j q^{k}   p_{4}^l ) 
	& = & \nonumber \\
{\rm sign}( \epsilon_{ijkl} p_{1}^i p_{2}^j p_{3}^k q^{l}   ) & = & 
	{\rm orientation} \, .
\end{eqnarray} 
In practice, at single precision there is a slight risk that $q$ will lie
on the boundary of one of the tetrahedra, and one of the above quantities
will be zero.  To cover this possibility we measure the winding number
using more than one $q$ and check that they agree.  In practice errors
are very rare, but we retain the checking feature as a precaution.

\subsection{gauge dependence and gauge choice}

Next let us consider the behavior of this winding number.  Suppose that
the slave field is slowly varying, ie its value at any pair of vertices on
a tetrahedron are separated by, say, less 
than $\pi/2$ radians.  Then moving the value of $S$ at a
single point by some small amount (say, less than $\pi/2$
radians) will cause the images of some tetrahedra to grow, at the expense
of others; but the winding number will not change.  To change winding
number we must have one of the tetrahedra expand until it covers half
the 3-sphere, at which point the geodesic rule switches which geodesic
tetrahedron on the 3-sphere it covers, and the winding number changes
by $\pm 1$.  This will only occur if the values of the slave field at
two neighboring points differ by a large
amount. Hence we find that, {\it if} the
initial slave field configuration is slowly varying, then the winding
number will be invariant to {\it small} gauge transformations, meaning
ones where the gauge element selected at each point is in the vicinity
of the origin.  This is the lattice analog of the invariance
of the continuum
winding number to small gauge transformations.  Similarly, if
the slave field is slowly varying, and we apply a large but slowly varying
gauge transformation (ie $g$ need not be in a close neighborhood of the
origin, but the angle between its values at neighboring points is always
small), the winding number will change by the winding 
number of the gauge transformation.  Note that both small gauge 
transformations and large but smooth gauge transformations 
may reduce the smoothness of the slave field, so a series of small
gauge transformations can eventually change the winding number.  Hence, 
the slave field winding number has properties much like what we need, 
but only when the slave field is kept slowly varying.

In general there will be two situations in which the slave field will not
be slowly varying.  One is avoidable, and the other is not. The first
occurs if there are link matrices which are very
far from the identity, in which case the slave fields on the two ends
of the link will want to differ from each other by an angle equal to
the angle by which the link differs from the identity.  Since
we are in three space dimensions, the thermodynamics of classical 
Yang-Mills theory is super-renormalizable, and if our lattice
spacing is small then the elementary plaquettes should all be close to the 
identity; so there should be some gauge which can prevent this from
happening.  We will discuss finding and staying in such a gauge below.
The other circumstance in which the slave field varies rapidly in space 
is when the links are all close to the identity but
there is a texture unwinding event in its final stages, i.e. a topological
winding in the slave field has been concentrated into a very small volume.
In the continuum such a winding is removed when the slave field develops
a singularity; on the lattice what happens is that the winding reaches
the lattice scale and then `slips through the lattice', leading
to a winding number change.  This is our signal that the gauge field
configuration has changed from being closer to one winding number vacuum
to being closer to another; in this case we must allow the slave field
to change winding number, and settle to the new $N_{CS}$ vacuum,
before performing a large gauge transformation to make $S(x)$ uniform again.


{} From this discussion, we see that
the slave field winding number will track $N_{CS}$ provided that
we can find a gauge which is smooth (meaning all link matrices are 
rather close to the identity) at any moment\footnote{Note that this
definition of smooth is very weak; in the continuum limit it does not
require that the fields be differentiable or even continuous, but only that
as the lattice spacing $a$ is made smaller, that $A$ grows slower than
as $a^{-1}$.
For classical fields, $A \sim a^{-1/2}$ at least in some gauge,
and the connection becomes rapidly smoother, in the sense used here, as
the lattice spacing is made smaller.}
and which does not
change abruptly between timesteps (the equivalent of the requirement in
the continuum case that the gauge is nonsingular).  Temporal gauge 
(time links set to the identity) satisfies the latter requirement; the 
gauge field evolves smoothly in time because in this gauge\footnote{
Here and throughout our notation is the same as \protect{\cite{AmbKras}}.
Our implementation of the equations of motion is that of
\protect{\cite{Ambjornetal}}, which differs at $O((\delta t)^3)$ from
what is written here.}
\begin{equation}
U_i(x,t + \delta t) = \exp( i \delta t \: \tau \cdot E_i(x, t + 
\delta t/2)) U_i(x,t) \, ,
\end{equation}
and $\delta t$ should be small (and so generally is $E$, which 
satisfies $E^2 \sim 1/\beta_L$).
It is also possible to choose a gauge for the initial conditions in which
the gauge field is smooth, by using the freedom in temporal gauge to
make a time independent gauge transformation.  One can take for instance
the lattice
equivalent of Coulomb gauge \cite{Mandula}, namely the gauge which minimizes
\begin{equation}
\sum_{x,i} \left( 1 - \frac{1}{2} {\rm Tr} U_i(x) \right) \, .
\end{equation}
This ``lattice Coulomb'' gauge should make the connection as smooth
as possible.  It is also the gauge in which the minimal energy slave
field configuration is everywhere the identity.  This gives a simple 
algorithm for finding this smoothest gauge; evolve the slave field
dissipatively to an energy minimum, and then gauge transform by 
$g(x) = S^{\dagger}(x)$.  Provided that the slave field was not in
the middle of a texture collapse when we terminated the dissipative
evolution, this should make the gauge connection smooth everywhere.
And there are simple gauge invariant measurements which can tell if
the slave field is close to a texture collapse event or some other 
phenomenon which will make the connections non-smooth in some small 
neighborhood.  One can define a peak stress, for instance,
\begin{equation}
{\rm peak \; stress} = \sup_{x}
	\left[ \sum_i \left( 2 - \frac{1}{2} {\rm Tr} (S^{\dagger}(x) 
	U_i(x) S(x+i)) - \frac{1}{2} {\rm Tr} (S^{\dagger}(x)
	U^{\dagger}_i(x-i) S(x-i)) \right) \right] \, .
\label{secondstress}
\end{equation}
If the peak stress exceeds some threshold then gauging to $S(x)=I$
will make the connection not smooth in some small neighborhood; 
otherwise the gauge $S(x) = I$ is smooth everywhere.

\begin{figure}
\centerline{\psfig{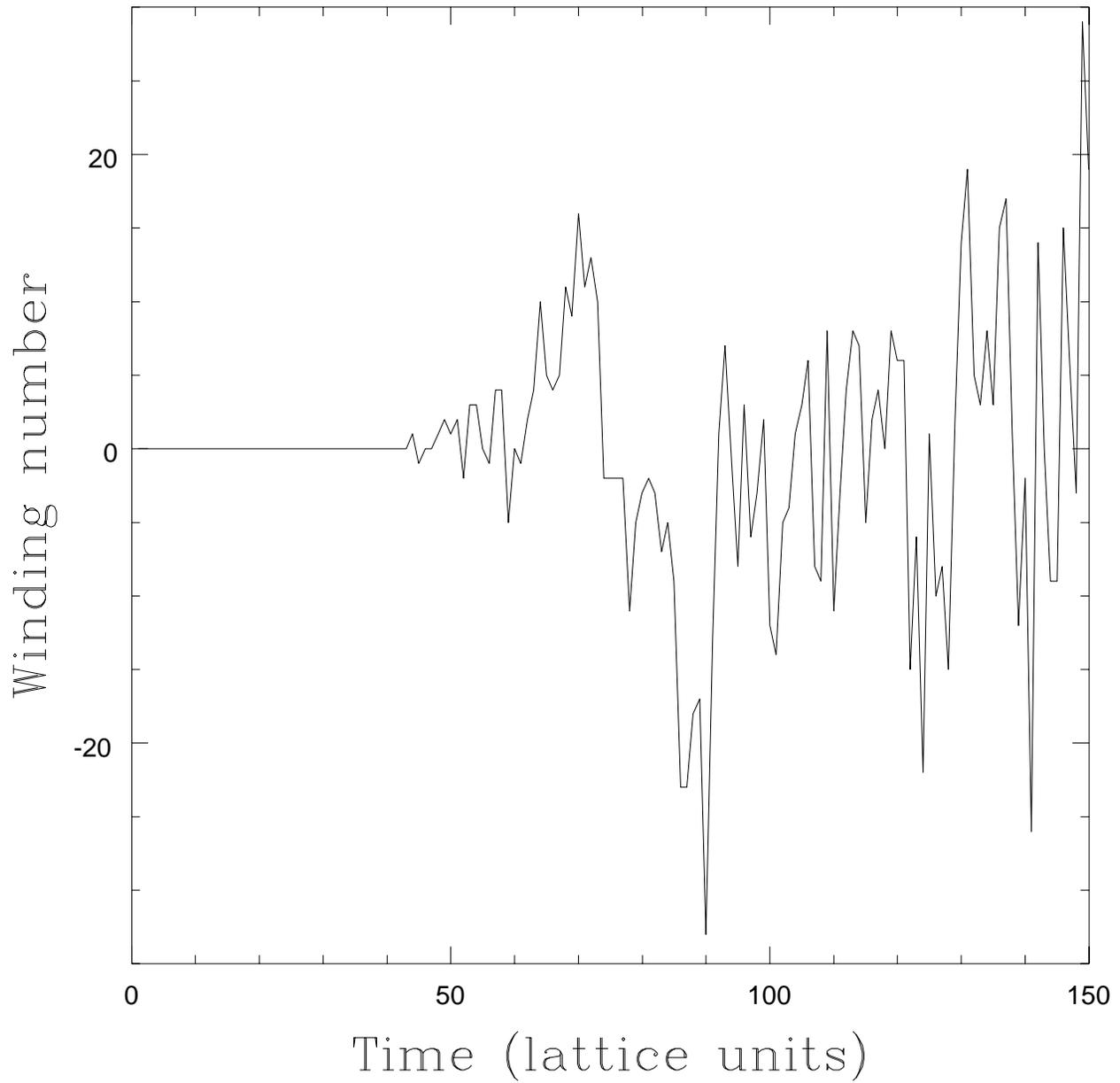}}
\caption{\label{nogauge} Evolution of winding number in the broken phase
of Yang-Mills Higgs theory, starting with smooth
connections and remaining in temporal gauge without ever performing
small gauge changes to keep the connections smooth.  
The connections gradually become less smooth, so after a
while the winding number begins to oscillate wildly.}
\end{figure}

Evolving the system in temporal gauge indefinitely 
is insufficient, however, because
the connections will gradually move away from being smooth.  The 
condition $\vec{\nabla} \cdot \vec{A} = 0$ is preserved by the evolution
in temporal gauge only if $\vec{\nabla} \cdot \vec{E}=0$; but
Gauss's law is $\vec{D} \cdot \vec{E} = 0$, which only agrees at leading
order.  Eventually the connections will become far from
smooth and the good behavior of the slave field winding number will 
break down.  This is illustrated in Figure \ref{nogauge}, which shows
the evolution of the slave field winding number in the broken phase of
Yang-Mills Higgs theory, on a $24^3$ lattice at $\beta_L = 8$,
starting in the gauge $S(x) = I$ and evolving the slave field with the
efficient dissipative dynamics discussed below.  The slave field
winding number remains zero, as it should, for some time, but the
connections and the minimum energy slave field configuration become ever 
less smooth, until the slave field winding number begins to oscillate
wildly.

Instead it is necessary to make occasional gauge changes to restore
the smoothness of the connection.  As discussed earlier, to preserve the
correct behavior of the slave field winding number these transforms
must be either small or large but smooth, both in the sense defined
earlier; and if the gauge transform is large, i.e. it changes the slave
field winding number, then we must know the winding number before and
after the gauge transformation, and add the change to a total 
count of windings removed by gauge
changes.  Our surrogate for $-N_{CS}$ is the current winding number
plus the count of windings removed by past gauge changes. 

If the (temporal gauge) evolution since the last gauge change has been 
relatively short and the slave field's peak stress was not large at the
beginning of, during, or at the end of the evolution, then the gauge
change to the gauge $S(x) = I$ will be small.  
If the slave field's peak stress was small at
the beginning and end of the evolution but not in between,
then the gauge change may be
large, but it should be smooth (since the connection is smooth before
and after).
If on the other hand the slave field's peak stress is large at the end
of the brief period of evolution, then the gauge change would probably
not be smooth, and it will make the connections unsmooth in some small
neighborhood; it is possible that making such gauge changes will cause
us to lose information about winding number change.  (For instance, if
we applied a gauge change to $S(x) = I$ at every timestep, then unless
the dissipative dynamics could fully collapse a texture event in a
single timestep, we would never observe any change in its winding number;
this is an example of a series of small gauge changes which change the
winding number, because they bring the connections through non-smooth
configurations in the process.)  So our algorithm is to measure
the slave field winding number every few (say 5) timesteps (remember that
a time step is $\delta t$ times a lattice unit of time evolution;
in all work in this paper $\delta t = 0.05$),
and provided that its peak stress falls below a threshold, we should
apply a gauge transformation to the gauge $S(x) = I$ everywhere; but if
the peak stress exceeds the threshold we should not change gauge, but 
should check again next time.  

We have computed that, for our
definition of peak stress, Eq. (\ref{secondstress}), if the plaquettes
are all close to the identity, then the peak stress during a winding
number changing evolution 
must exceed 2.0 at some time, and if we gauge change just before
it first reaches 1.0 and again after it falls back to 1.0 we will always
correctly identify the winding number change.  These estimates are 
very conservative, and in practice, with smooth connections, the stress 
always peaks at at least 2.4 during texture collapse, and generally much
higher (depending on how close the core of the texture is to a lattice
site); and for the real case the ultraviolet fluctuations in the magnetic
field also contribute to the peak stress, 
and the peak value is still higher.  We have used
a threshold of 1.2 in our work.  We have
verified for long evolutions that changing the 
threshold to 1.1 or 1.3 does not change the determined winding
number, although it is also clear that 
a sufficiently high or low threshold would either allow
true winding number changes to be gauged away, or prevent smoothing
gauge changes for long enough that the configuration becomes non-smooth,
in which case the winding number is sensitive to small changes in the
configuration and loses its topological interpretation.

\subsection{dissipative evolution algorithm}

All that remains is to specify an efficient dissipative algorithm for
keeping the slave field near the minimal energy configuration while the
gauge fields evolve.  An element of a simple quench algorithm 
developed by Mandula and Ogilvie \cite{Mandula} is to 
minimize the Hamiltonian with respect to variations in the slave field
at one point, $S(x)$.  That is, 
we replace $S(x)$ with the value which minimizes
\begin{equation}
\sum_i \left( 2 - \frac{1}{2} {\rm Tr} (S^{\dagger}(x) 
	U^{\dagger}_i(x-i) S(x-i)) 
	-  \frac{1}{2} {\rm Tr} (S^{\dagger}(x) 
	U_i(x) S(x+i)) \right) \, ,
\end{equation}
which is minimized by choosing $S(x)$ to be the sum of the parallel 
transports of the nearest neighbors, treated as points in $\Re^4$,
projected to unit modulus.  (Note that the length of the vector in $\Re^4$
before the projection is 6 minus the ``stress'' at
that point, which is then a biproduct of the calculation.)
A quench of the slave field consists of sweeping over lattice 
sites in some order, applying this algorithm at
each.  The updates of even and odd sites are independent,
so it is easiest to alternately update all even and odd sites. 
A weak sinusoidal perturbation to $S(x)$, with
wave number $k$ and lattice dispersion measure $\omega^2 = \sum_i
4 \sin^2(k_i/2)$, will be suppressed by a factor of $(1-\omega^2/12)^2$
for each update of all sites, so the quench is very efficient at removing
high $k$ excitations but less effective in the infrared.

The simplest quench algorithm is to first take $S(x,t+\delta t) = S(x,t)$
and then apply the above quench once (or several times) each timestep.
This can be significantly improved by using ``memory'', or giving the
system damped inertial dynamics.  Since the timestep is much shorter than
any inverse frequency in the system's dynamics, the update of $S$ at
one timestep will be almost the same as for the previous one; so we 
apply $m<1$ times the last time's update pre-emptively before quenching.
In other words, rather than taking $S(x,t+\delta t) = S(x,t)$ and 
then quenching, we take $S(x,t+\delta t) = (S(x,t) S^{\dagger}(x,t - 
\delta t))^{m} S(x,t)$ as our initial guess and then quench.  Here 
$S$ at previous timesteps is always the value after the quench was
applied at that timestep.  The quantity $S(x,t) S^{\dagger}(x,t - 
\delta t)$ can be viewed as the slave field's momentum and is stored
as a Lie algebra element (so taking the $m$ power becomes multiplication
by $m$).  $(1-m)$ acts as a damping coeffient and should be positive
for stability; we find $1-m = \delta t$ proves quite efficient, 
increasing the performance of the algorithm, in the absence of a collapse
event in its final stages, by on order $1/\delta t$.

Two further modifications are wise, because of the ``texture collapse''
events.  In such an event the slave field changes very rapidly
as it removes the unwanted winding, after which the direction it needs
to evolve abruptly changes.  We prevent rebounding after the
collapse event by turning down or off the memory effect at sites 
where $S(x,t) 
S^{\dagger}(x,t - \delta t)$ is too far from the identity,
say $1 - (1/2) {\rm Tr} S(x,t) S^{\dagger}(x,t - \delta t) > (1-m)^2$.
The slave field then has fast, damped inertial dynamics with nonlinear
damping coefficient.  Also, to make the final collapse of the textures
more efficient (and to minimize the time spent with the peak stress
above the threshold), whenever the peak stress goes above its
threshold we triple the number of quenches used, and substitute $m^3$
for $m$, in a neighborhood of the most stressed point.

We have run several tests of this quench algorithm.  We have evolved two
slave fields simultaneously on the same gauge field background, taken
from a real simulation, single quenching one and double quenching the
other; we find almost no difference between how well they minimize the
slave field
energy, and the double quenched slave field typically discovers winding
number changes only slightly sooner (at most 1 lattice length of time, or
20 updates) than the single quenched slave field.  We also evolved two
slave fields simultaneously, each single quenched, but beginning one of 
them from the minimal energy configuration and the other from a completely
random initial configuration.  Within one lattice unit of time their
energies were comparable and their winding numbers differed by at most 2,
and after 5 lattice units of time their winding numbers ceased ever to 
differ and they became essentially identical.

\section{Numerical results}
\label{four}

Here we present results of numerical investigations of the motion of
$N_{CS}$ using the slave field technique.

\subsection{tests} 

First we tested the slave field technique on ``mocked up'' backgrounds to
see that it behaves as expected.  In the naive vacuum, starting the slave
field in a winding 1 configuration and applying the dissipative algorithm,
we observe the slave field energy to
drop while the peak stress increases, peaking at a
value always above 2.4, at which point the winding number changes to zero.
The energy and peak stress then fall off.  This is the behavior expected
for a ``texture collapse'' event.  We have also ``mocked up'' a Sphaleron
transition; we find an $N_{CS}=1$ vacuum 
gauge field configuration by gauging away
a winding one slave field configuration with initially flat connection;
we then take the $n$'th root of each gauge link and ``evolve to'' 
the winding 1 gauge field configuration by starting with flat connections
and repeatedly multiplying each link by the appropriate $n$'th root.
The slave field evolves dissipatively in this background.  It changes
winding number slightly more than halfway through the evolution, shortly
after the underlying configuration switches from being nearer one vacuum
configuration to being nearer the other.  As we conduct the
evolution in more steps, the 
slave field changes winding number more steps, but a smaller fraction of
the length of the evolution, after the midpoint of the evolution.  The
response of the slave field to a change in the winding number of the
underlying configuration is delayed, but not very much; the sluggishness
to respond to winding number change will translate in a real simulation
with a timestep $\delta t = 0.05$ to a delay of about one lattice unit
of time, which is small.  We also observe that the ``old'' definition of
Chern-Simons number systematically underestimates the actual $N_{CS}$ change
through this evolution, by an amount which grows worse as the initial winding 
becomes less spread out.  This is the result of nonrenormalizable operator
corrections and we will discuss it again below.

\subsection{comparison with the cooled field technique}

Next, we want to see if the slave field method will be reliable for tracking
$N_{CS}$ in real simulations of Yang-Mills or Yang-Mills Higgs theory.
It would be illuminating to compare it against the ``old'' definition, Eq. 
(\ref{olddef}), except that we strongly believe that this definition is
contaminated by lattice artifact diffusion.  However, Ambj{\o}rn and 
Krasnitz have recently proposed a patch for this definition which should
eliminate almost all of its white noise and lattice artifact diffusion
\cite{AmbKras2}.  The idea is to copy the connections at each
timestep and quench or ``cool'' them, and then to track $N_{CS}$ for the
evolution of these cooled fields.  The infrared fields, which are responsible
for the diffusion of $N_{CS}$, will be unaffected, but the ultraviolet 
excitations on top of them will be removed, and will no longer contribute
white noise, lattice artifact diffusion, or screening of the field operators
in the old definition of $N_{CS}$, Eq. (\ref{olddef}).  The cooling is 
just an evolution of fixed length under straight dissipative dynamics, 
$\dot{U} = - \partial H/\partial U$ (correctly interpreted for group 
elements).  

\begin{figure}
\centerline{\psfig{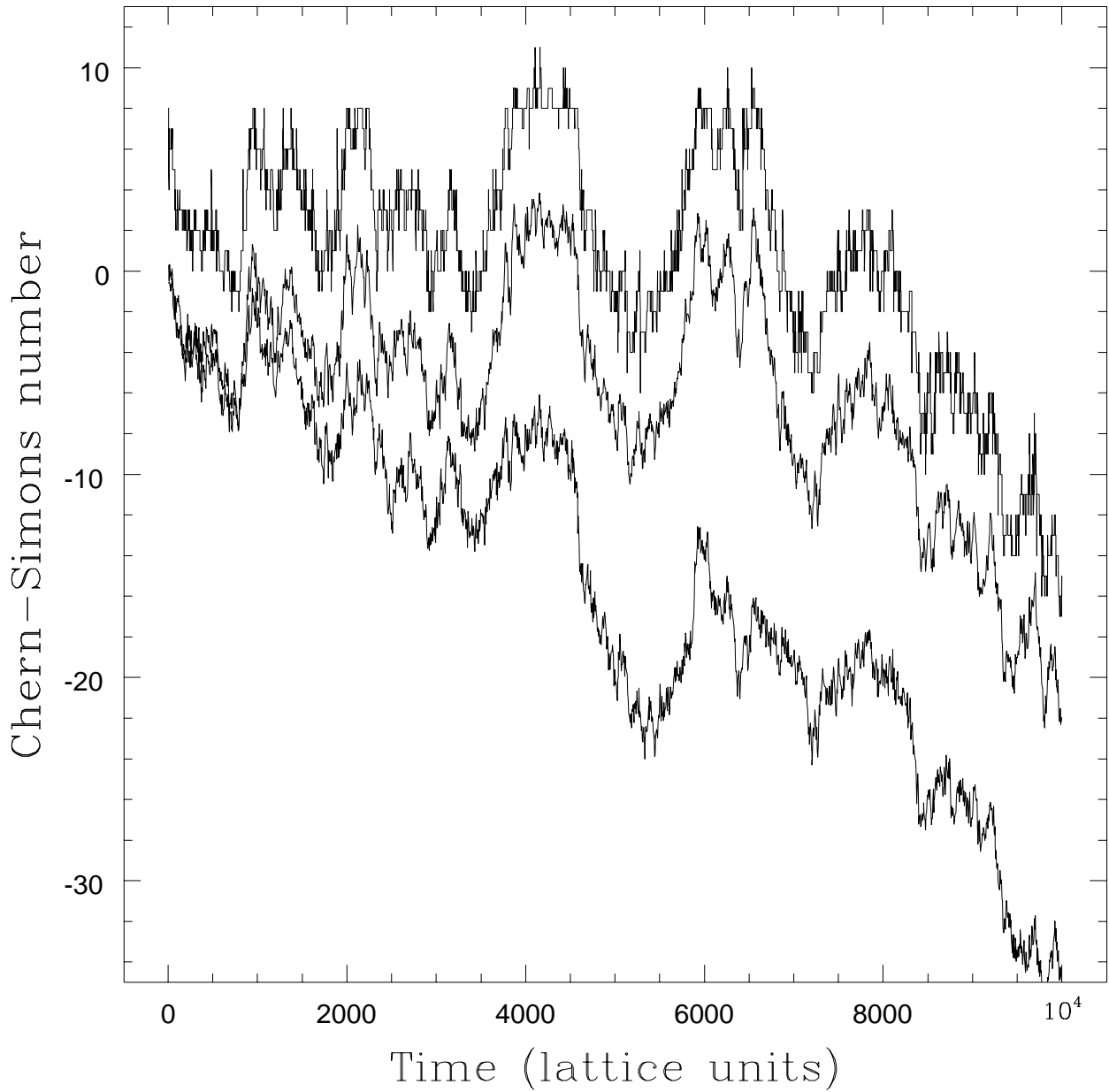}}
\caption{\label{3NCS} Chern-Simons number, tracked three ways, as a function
of time, each for the same Hamiltonian evolution.  The stairstep curve (top)
is the winding number definition, the curve which tracks it closely is
the cooled field definition, and the curve which
drifts away from them is the old definition.  The winding number definition
has been shifted for clarity, otherwise it would land almost exactly on
the cooled field definition.  These two are very highly correlated, and
the ``old'' definition is quite correlated with them but contains in
addition some extra diffusive signal.}
\end{figure}

We implemented this algorithm with a cooling of length 
of 1.25 lattice units (which should be enough to remove the worst
behaved frequencies and greatly ameliorate the problems with
the old definition, although a longer cooling might be preferable).
We then evolved Yang-Mills Higgs theory in the symmetric phase, just at
the phase transition temperature, recording $N_{CS}$ using all three 
definitions; the results are presented in Figure \ref{3NCS}.  These data
were taken on a $24^3$ grid at $\beta_L = 8.07$, $\lambda_L = 0.2$,
and $m_{HL}^2 = -0.3223$ (bare), the same values generally
used in \cite{MooreTurok}; we also use the thermalization and evolution
algorithms and the definition of temperature used there.  The total
length of the evolution was 10500 lattice units.  The two
new definitions of $N_{CS}$,
the slave field definition and the cooled field definition,
agree very closely, though there is considerable high frequency noise.
Some of this noise might arise because the cooling step in the cooled field
approach changes $N_{CS}$, but probably most of it is due to the problems
with the slave field definition which were discussed at the end of
Section \ref{two}.

To study how well these methods agree and what diffusion constant they
imply, we apply a cosine transform and study their frequency 
spectra\footnote{In \protect{\cite{MooreTurok}} we advocated using a 
sine transform.  That approach is not time symmetric, and when white
noise is present, the noise in the first datapoint is propagated into
all the data.  The cosine transform used here gives correct performance
in the presence of white noise.}.
Writing $N_{CS}$ as a function of time as $z(t)$, we define
\begin{equation}
\tilde{z}(m) = \int_0^{t_f} \frac{dt}{t_f} z(t) \cos \left( \frac{ m \pi t}
	{t_f} \right) \, ,
\end{equation}
and find that, if $z(t)$ evolves diffusively with diffusion constant
$\langle ( z(t) - z(t'))^2 \rangle = \Gamma | t - t' |$, then $\tilde{z}(m)$
will be Gaussian distributed and independent, with variance
\begin{equation}
\langle \tilde{z}(m) \tilde{z}(n) \rangle = \frac{\Gamma t_f}{2 \pi^2 m^2}
	\delta_{mn} \, .
\label{fourierspectrum}
\end{equation}
When the sampling is discrete with $N$ intervals between measurements, then
the integral should be replaced by a sum, with $z$ at each endpoint only
sampled with half weight; in this case $m$ on the rhs. 
of Eq. (\ref{fourierspectrum}) should be replaced
with $(2N/\pi) \sin(m \pi/2N)$.  The addition of white noise will add an
$m$ independent term, but the transform
coefficients remain independent and Gaussian, which makes this decomposition
particularly convenient for analysis.  Other sorts of noise could make
the Fourier components nonGaussian or correlated, but this
will not be a problem if we only use the most infrared Fourier modes in
the analysis, since on long time scales the evolution should be almost
purely diffusive.

We use this technique to compare the slave field, cooled field, and 
``old'' definitions of $N_{CS}$.  Denote their cosine transform spectra
as $\tilde{z}_{s}$ , $\tilde{z}_{c}$, and $\tilde{z}_{o}$ respectively.  
We will be concerned
only with the infrared, since we want to know how they track $N_{CS}$
over long times
and if there are noise contributions to its diffusion constant, so
we will consider sums over the first several terms, reweighted by multiplying
each $z$ by $m$ so as not to overemphasize the first few coefficients.

First, let us assume that the cooled field reports only $N_{CS}$ but that
it has a multiplicative renormalization, because of nonrenormalizable
operators and screening from the modes which were not completely quenched.
The slave field method, being topological, should track $N_{CS}$ 
unrenormalized, but perhaps with some extra diffusive signal if there are
occasional erroneous identifications of winding number change.  We can find
the multiplicative renormalization of the cooled field technique by
measuring $\tilde{z}_{c} \cdot \tilde{z}_{s} / \tilde{z}_{c}^2$,
where
\begin{equation}
\tilde{z}_1 \cdot \tilde{z}_2 \equiv \sum_{n=1}^{N} m^2 
	\tilde{z}_{1}(m) \tilde{z}_{2}(m)  \, .
\end{equation}
The noise part of $\tilde{z}_{s}$ is uncorrelated and will not 
contribute (on average), and we will just
find the inverse of the renormalization of the cooled field 
definition.  The result, using the first 25 to 50 transform coefficients,
is $\tilde{z}_{c} \cdot \tilde{z}_{s} / \tilde{z}_{c}^2 = 
1.05 \pm 0.01$ (the error bar reflecting dependence on the number
of points used).  Hence, the cooled field technique underreports winding
number change by a factor of $0.95$, for this level of cooling and 
this lattice spacing\footnote{This under-reporting could be ameliorated by
cooling more heavily and using an improved local operator free of $O(a^2)$
errors for $E \cdot B$, and by going to a finer lattice.  We do not view 
it as a fundamental defect of the cooled field method.}.

\begin{figure}
\centerline{\psfig{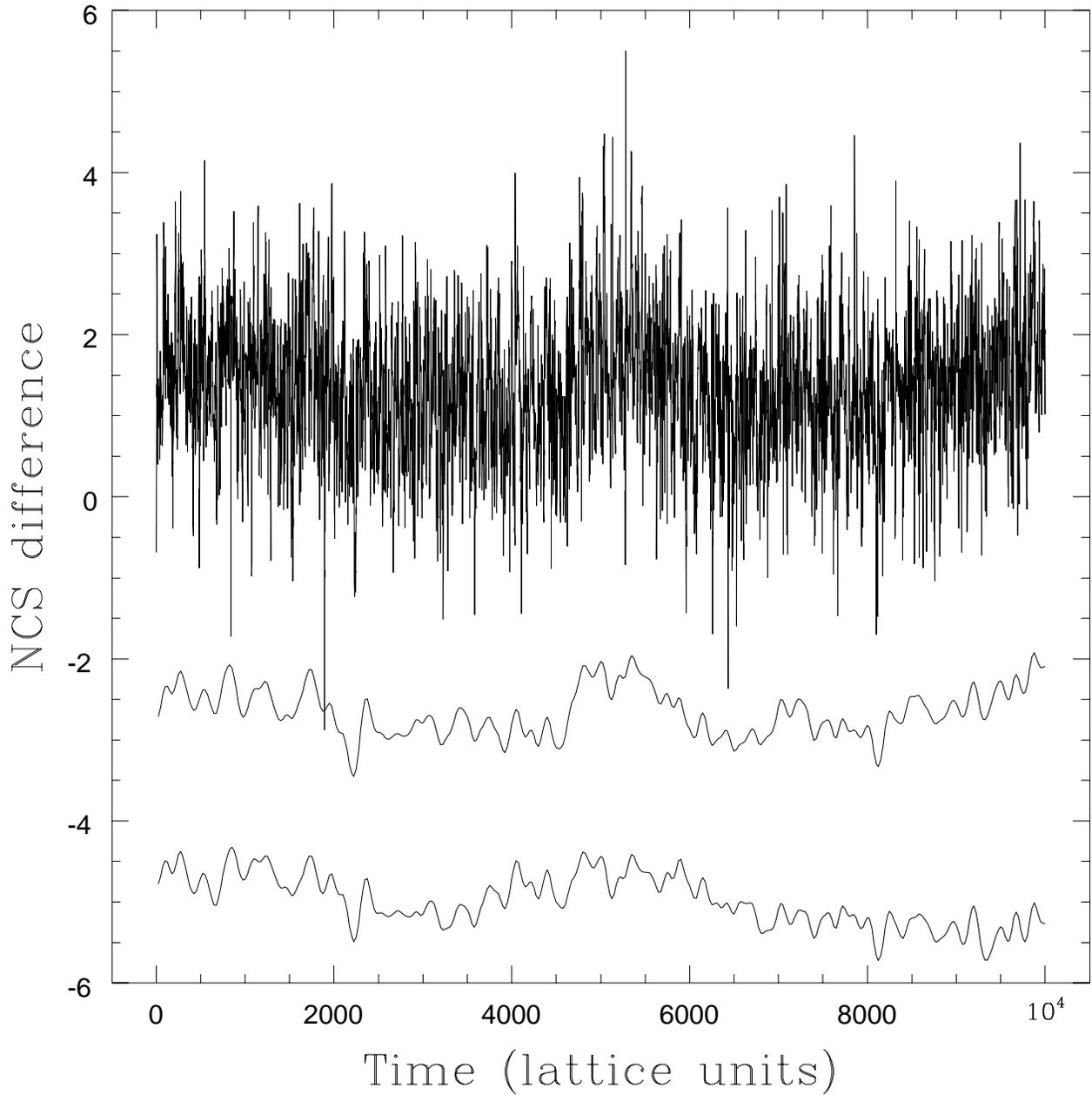}}
\caption{\label{offby1} The difference between the winding number definition
of $N_{CS}$ and the cooled field definition (top curve), 
with moving average (shifted for clarity, middle curve), and the moving
average after correlations with the cooled field definition have been
removed (bottom curve).  The correlations mean that the cooled field
receives a multiplicative correction with respect to the winding number
(topological) definition, probably due to nonrenormalizable operators.}
\end{figure}

We can also see how large the noise in the slave field
definition is from $1 - (\tilde{z}_{c} \cdot \tilde{z}_{s})^2 / 
( \tilde{z}_{c}^2 \tilde{z}_{s}^2)$, which measures the diffusive power
in the slave field definition which is not correlated with the cooled
field definition (and must be noise, in one definition or the other).
The value, using $N=25$, was $0.002$;
except for the
multiplicative renormalization of the cooled field technique, the two
methods agree in the infrared almost exactly.  In fact, the difference
is too small to be explained even by one accidental miscounting of a
winding number change by the slave field technique, as can be seen in
Figure \ref{offby1}.   The difference between the definitions is larger
at higher frequencies; if we had used a much larger value of $N$ above
we would have found poorer agreement.  This represents
non-diffusive noise in the slave field definition, 
presumably due to the algorithm
picking Gribov copies with different winding number from the
one which absolutely minimizes the slave field energy.  We can see from
Figure \ref{offby1} that the slave field winding number 
is often off by $\pm 2$ from the cooled field value, but never for long,
and the moving averages are in excellent agreement.
Also note that it is possible that the cooled field technique
has some residual spurious diffusion, because the ultraviolet modes are
not completely cooled; but the quality of the agreement between the
slave and cooled field methods in the infrared indicates that even with
a relatively modest cooling depth of $1.25$ lattice units,
spurious UV contributions to the diffusion constant for the smoothed
field definition are very small.

It is also instructive to apply these analyses to see how accurate the 
``old'' definition of $N_{CS}$ is, taking the slave field definition as
a benchmark.  for this run, using the first 40 transform coefficients,
we find $\tilde{z}_o \cdot \tilde{z}_{s}
/ \tilde{z}_s^2 = 0.67 \pm .06$ and 
$1 - (\tilde{z}_{o} \cdot \tilde{z}_{s})^2 / 
( \tilde{z}_{o}^2 \tilde{z}_{s}^2) = 0.28 \pm .06$.  The renormalization of
the old definition is startlingly large, and there is considerable 
diffusive power uncorrelated with true topology change.  
(The error bars are estimates based on the expected RMS ``accidental''
projection of the noise part of the old definition of $N_{CS}$ 
along the slave field definition, and the expected statistical fluctuations
in the size of the noise part, $1/\sqrt{N}$ of its amplitude.)

\begin{figure}
\centerline{\psfig{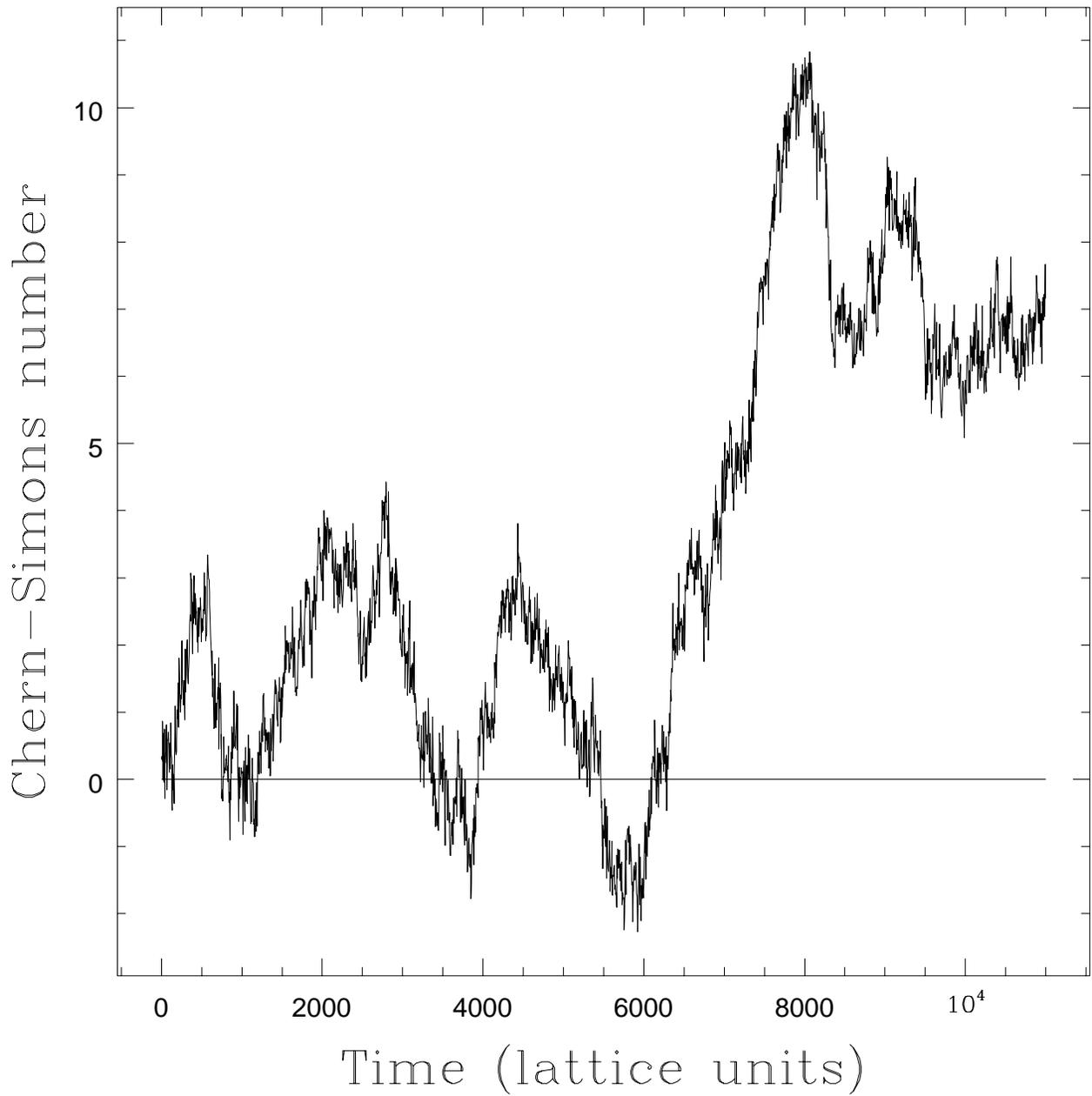}}
\caption{\label{brokenphase} winding number (constant line) and old definition
of $N_{CS}$ for a Hamiltonian trajectory in the broken electroweak phase
at the phase transition temperature.  There are no winding number changes
in the trajectory; the diffusion of $N_{CS}$ under the old definition is
a lattice artifact.}
\end{figure}

Another problem we can address with the slave field technique is the 
diffusion rate of $N_{CS}$ in the broken electroweak phase.  To do this
we cooled and then heated the starting configuration for the symmetric
phase run discussed above, to bring it into the broken electroweak phase
at exactly the same temperature.  We then made another Hamiltonian 
evolution of $11000$ lattice units length.  The slave field winding number
and the old definition of $N_{CS}$ for this run are presented in Figure 
\ref{brokenphase}; while the old definition of $N_{CS}$ indicates 
considerable diffusion, the slave field technique shows that not a 
single winding number change occurred.  This confirms the conjecture
\cite{MooreTurok} that the diffusion observed in the broken electroweak
phase using the old definition of $N_{CS}$ is a lattice artifact, and
it also agrees with the expectations from the semiclassical ``Sphaleron''
calculation that the rate should be exponentially small 
\cite{ArnoldMcLerran}.  In particular this run bounds the diffusion constant
for these values of parameters ( the broken phase here has $\phi = 1.4
gT$) to be $\kappa < 0.002$, though the error bar here is not Gaussian.

\subsection{lattice spacing dependence}

Finally we will investigate how the diffusion constant for $N_{CS}$
scales with lattice spacing.  In Yang-Mills Higgs theory in the symmetric
phase the rate will depend weakly on the distance from the phase transition
temperature and on the scalar self-coupling.  The physically interesting
diffusion rate will be the rate in Yang-Mills Higgs theory in the 
symmetric phase with some supercooling, at a value of the scalar 
self-coupling which is not yet known.  However, we are interested here only
in the lattice spacing dependence of the rate and these other problems
only complicate matters, so we analyze pure SU(2) Yang-Mills theory.

Before presenting the results we should point out that even if the 
rate of $N_{CS}$ diffusion is independent of the strength of charge 
screening, we should still expect a weak lattice spacing dependence 
in our results if we translate the diffusion rate per unit volume in
lattice units into the diffusion rate per unit physical volume using
the tree level relation $\beta_L = 4/(g^2 a T)$.  As discussed in 
\cite{Oapaper}, the wave function normalization on the lattice will
receive $O(\beta_L^{-1})$ corrections because the ultraviolet modes
screen the infrared physics differently on the lattice than in the
continuum; these corrections can be understood as a correction in the 
matching between the lattice length scale and the physical length 
scale.  There is also a correction to the infrared thermodynamics
from the $A_0$ fields which depends on the Debye mass, which varies
with lattice spacing in SU(2) Yang-Mills theory with the Kogut-Susskind
Hamiltonian used here as
\begin{equation}
m_D^2 = \frac{4 g^2 \Sigma T}{4 \pi a} = \frac{\Sigma \beta_L}{4 \pi} 
	g^4 T^2 \, .
\end{equation}

The former correction can be extracted from results in \cite{Oapaper}
by setting $\tan \Theta_W = 0$, dropping all scalar contributions,
and using the results in Appendix C of that paper to include the $A_0$ field;
we find that the lattice volume should be converted into physical 
volume by using\footnote{Note that an early preprint version of 
\protect{\cite{Oapaper}} contained an algebraic error, producing a slightly
different expression.  Here we use the corrected version of that paper.}
\begin{equation}
\beta_{L,{\rm imp}} = \beta_L - \left[ \frac{70}{3} \frac{ \xi}{4 \pi} + 
\frac{1}{3} \frac{\Sigma}{4 \pi} + \frac{1}{3} \right]
= \beta_L - 0.701  \qquad (\xi = 0.152859325)
\end{equation}
in place of $\beta_L$ in the formula for $a$.  This gives the correct
relation to a three dimensional continuum theory which, however, has a wrong 
(and lattice spacing dependent) Debye mass.  To correct for the different
Debye mass it is best to express the final answer in terms of $\bar{g}_3^2$,
the natural inverse length scale of the 3 dimensional theory with the 
$A_0$ field integrated out. $\bar{g}_3^2$ sets the scale for nonperturbative 
infrared physics.  It is related to $g^2$ by \cite{FKRS}
\begin{equation}
\bar{g}^2_3 = g^2 T \left( 1 - \frac{g^2 T}{24 \pi m_D} \right) \, .
\end{equation}

This settles how to minimize errors in converting spatial lengths 
between the lattice and the continuum; but the corrections to the 
rescaling of the time direction need not be the same.  This rescaling 
involves ultraviolet lattice artifacts in the real time 
evolution of the lattice system, which have not yet been studied in
the literature.  We will not solve this problem here, but we will
make an educated guess, based on reasoning presented in Appendix A, that
the $O(a)$ correction should be half as large as for a spatial direction.
The final relation between $\Gamma$, the diffusion constant for
$N_{CS}$ in lattice units, and $\bar{\kappa}$, the diffusion constant
in terms of the characteristic length scale $1/\bar{g}_3^2$ for 
nonperturbative physics, is then
\begin{equation}
\bar{\kappa} \equiv \Gamma({\rm phys. \; units}) 
	\left( \frac{4 \pi}{\bar{g}_3^2} \right)^4
	= \Gamma({\rm lattice \; units}) 
	(\pi \beta_{L})^4 \left( \frac{\beta_{L, {\rm imp}}}{\beta_L}
	\right)^{3.5} \left( 1 - \frac{1}
	{6 \sqrt{4 \pi \Sigma \beta_L}} \right)^{-4} \, .
\end{equation}
We will present our results in terms of this $\bar{\kappa}$, and as
$\kappa= \Gamma({\rm lattice \; units}) (\pi \beta_L)^4$ (without
thermodynamic corrections) for comparison with other literature.

We are interested here in finding the large volume limit of $\bar{\kappa}$
at several lattice spacings.  As we have seen, the old definition of
$N_{CS}$ is predominantly signal with a smaller, UV dominated noise
component; so studies of the volume dependence of the diffusion constant,
using the old definition, should reliably identify what a sufficient
volume to achieve the large volume limit is.  Ambj{\o}rn and Krasnitz used
this definition and found that when the lattice exceeds $2 \beta_L$
points on a side, the rate has achieved a large volume limit \cite{AmbKras}.
We use a lattice $3 \beta_L$ on a side, which should be abundantly 
sufficient, and we use this same physical volume for each lattice spacing 
to prevent any systematic differences between lattice spacings due to 
residual lattice volume dependence.  We have studied
lattices with $\beta_L = 6,8,10,12$, and 16.  

If $N_{CS}$ followed a perfect random walk then we could assume that 
the cosine transform coefficients $\tilde{z}(m)$ were Gaussian with
variance $\langle \tilde{z}^2(m) \rangle = A/m^2$
and convert $A$ into the diffusion constant.  But realistically the
motion of $N_{CS}$ will only be approximately diffusive on long time
scales, with corrections at finite $m$.  If we had a lot of data then we
could fit just the very low frequency coefficients to such a form, but
to get good statistics it is necessary to make a fit of a larger number
of coefficients with some assumption about the noise.  At the same time
we want to make sure that systematic errors in the fitting procedure are
smaller than statistical errors.  At finite $m$ we expect that
$\tilde{z}(m)$ should become weakly nonGaussian, that there should be
small cross-correlations between different $\tilde{z}(m)$, and that the
variance $\langle \tilde{z}^2(m) \rangle$ should receive corrections
from the $A/m^2$ form.  If we only study the values of $\tilde{z}^2(m)$
then only the latter needs accounting for to find the diffusion
coefficient.  The first nontrivial correction which can occur at small
$m$ is $\langle \tilde{z}^2(m) \rangle = A/m^2 + B$.  With this in mind
we fit the first several 
cosine transform coefficients to $m^2 \tilde{z}^2 = A + B m^2$.  
We vary the number
of coefficients used and find the point where the residuals first 
show a trend or the fit becomes poor, or the $B m^2$ term becomes
comparable to the $A$ term for the largest $m$; we then fit using half
this many coeffecients, to be sure that the fitting {\it Ansatz} is
still applicable.  
We have found that both $A$ and its error are weakly
dependent on where we place the cut, and we have also checked the
technique with blind tests on artificially generated (diffusion plus
noise) data.  Note that the {\it Ansatz} we use for the behavior of
$N_{CS}$ is that it is a Brownian signal plus noise which is white at
least on long time scales.  This is the same assumption made in the
fitting procedure used by Ambjorn and Krasnitz \cite{AmbKras,AmbKras2}.

\begin{table}
\centerline{\mbox{
\begin{tabular}{|c|c|c|c|c|c|} \hline
Phase & $\beta_L$ & lattice 4-volume & 
Def. of $N_{CS}$ & $\kappa$ & $\bar{\kappa}$ \\ \hline
broken & 8.07 & $24^3 \times 11000$ & old & 
$0.15 \pm 0.01$ & $0.11 \pm 0.01$ \\ \hline
broken & 8.07 & $24^3 \times 11000$ &  slave &
 $0.000 \pm 0.003$ & $0.000 \pm 0.002$ \\ \hline 
symmetric & 8.07 & $24^3 \times 10500$ &  old &
 $0.93 \pm 0.05$ &  $0.70 \pm 0.04$ \\ \hline
symmetric & 8.07 & $24^3 \times 10500$ &  cooled &
 $1.41 \pm 0.13$ &  $1.07 \pm 0.10$ \\ \hline
symmetric & 8.07 & $24^3 \times 10500$ &  slave &
 $1.57 \pm 0.14$ &  $1.19 \pm 0.11$ \\ \hline
Yang-Mills & 6 & $16^3 \times 32000$ &  slave &
 $2.61 \pm 0.18$ & $1.77 \pm 0.12$ \\ \hline
Yang-Mills & 8 & $24^3 \times 33000$ &  slave &
 $2.25 \pm 0.19$ & $1.70 \pm 0.14$ \\ \hline
Yang-Mills & 10 & $30^3 \times 50000$ &  slave &
 $1.91 \pm 0.12$ & $1.53 \pm 0.10$ \\ \hline
Yang-Mills & 12 & $36^3 \times 60000$ &  slave &
 $1.75 \pm 0.12$ & $1.46 \pm 0.10$ \\ \hline
Yang-Mills & 16 & $48^3 \times 75000$ &  slave &
 $1.34 \pm 0.10$ & $1.18 \pm 0.09$ \\ \hline
\end{tabular} }}
\caption{\label{resulttable} Results for $\kappa$ (naive lattice
continuum relation) and $\bar{\kappa}$ (relation including
thermodynamics corrections) in Yang-Mills
theory at several lattice spacings, and for Yang-Mills Higgs theory in
each phase and for 
three measurement techniques for $N_{CS}$.}
\end{table}

\begin{figure}
\centerline{\psfig{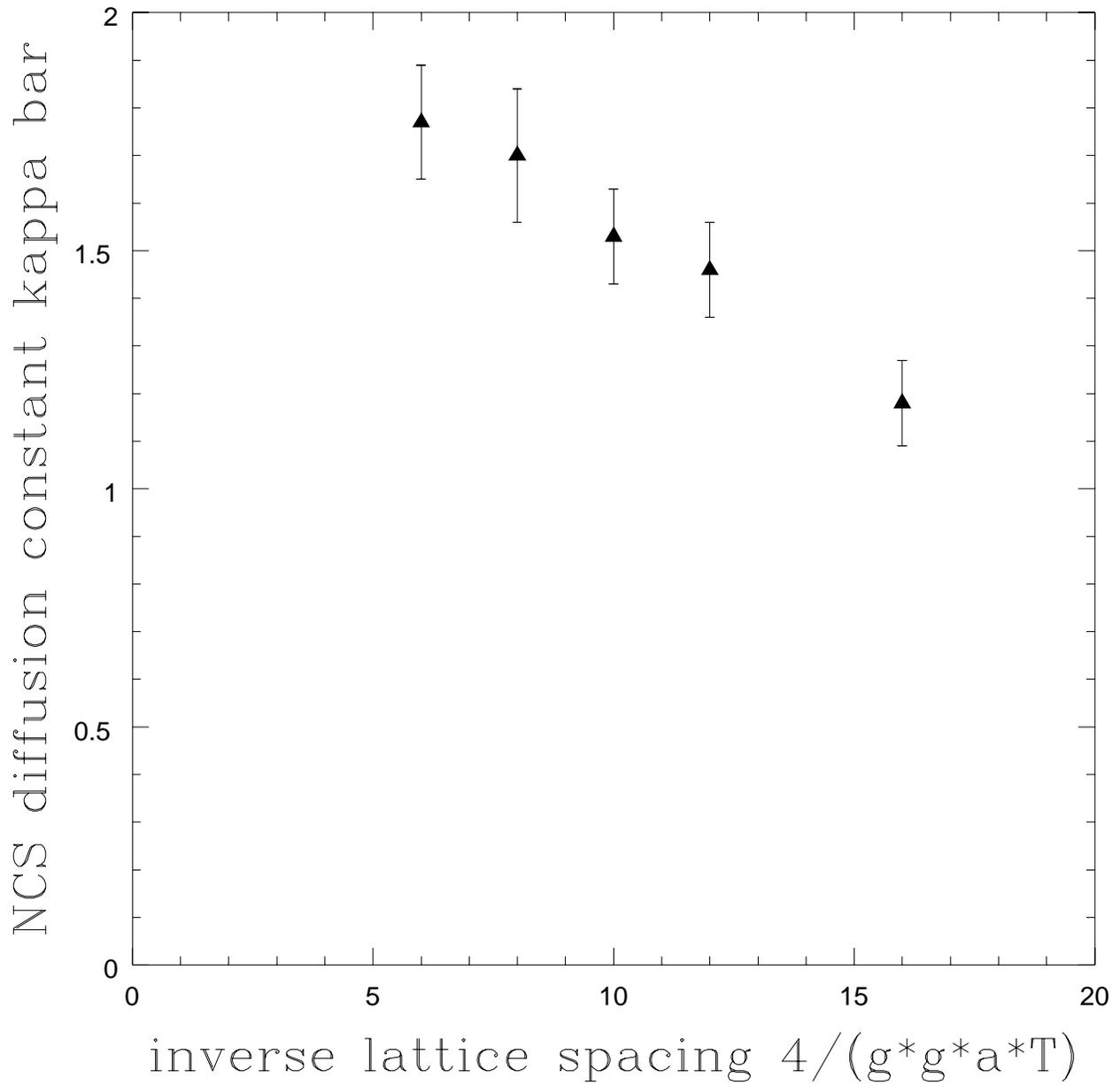}}
\caption{\label{resultfig} Results for $\bar{\kappa}$ as a function of
lattice spacing in Yang-Mills theory.  The dependence on $\beta_L$ is
strong, but not as strong as $\beta_L^{-1}$.}
\end{figure}

The results, presented
in Table \ref{resulttable} and Figure \ref{resultfig}, distinctly show
nonvanishing dependence on lattice spacing in the diffusion rate per
unit physical 4-volume.  However, the dependence is not as large as 
predicted by Arnold, Son, and Yaffe \cite{ArnoldYaffe}, who argue that
the data should be proportional to $\beta_L^{-1}$ (proportional
to $a$).  

\begin{figure}
\centerline{\psfig{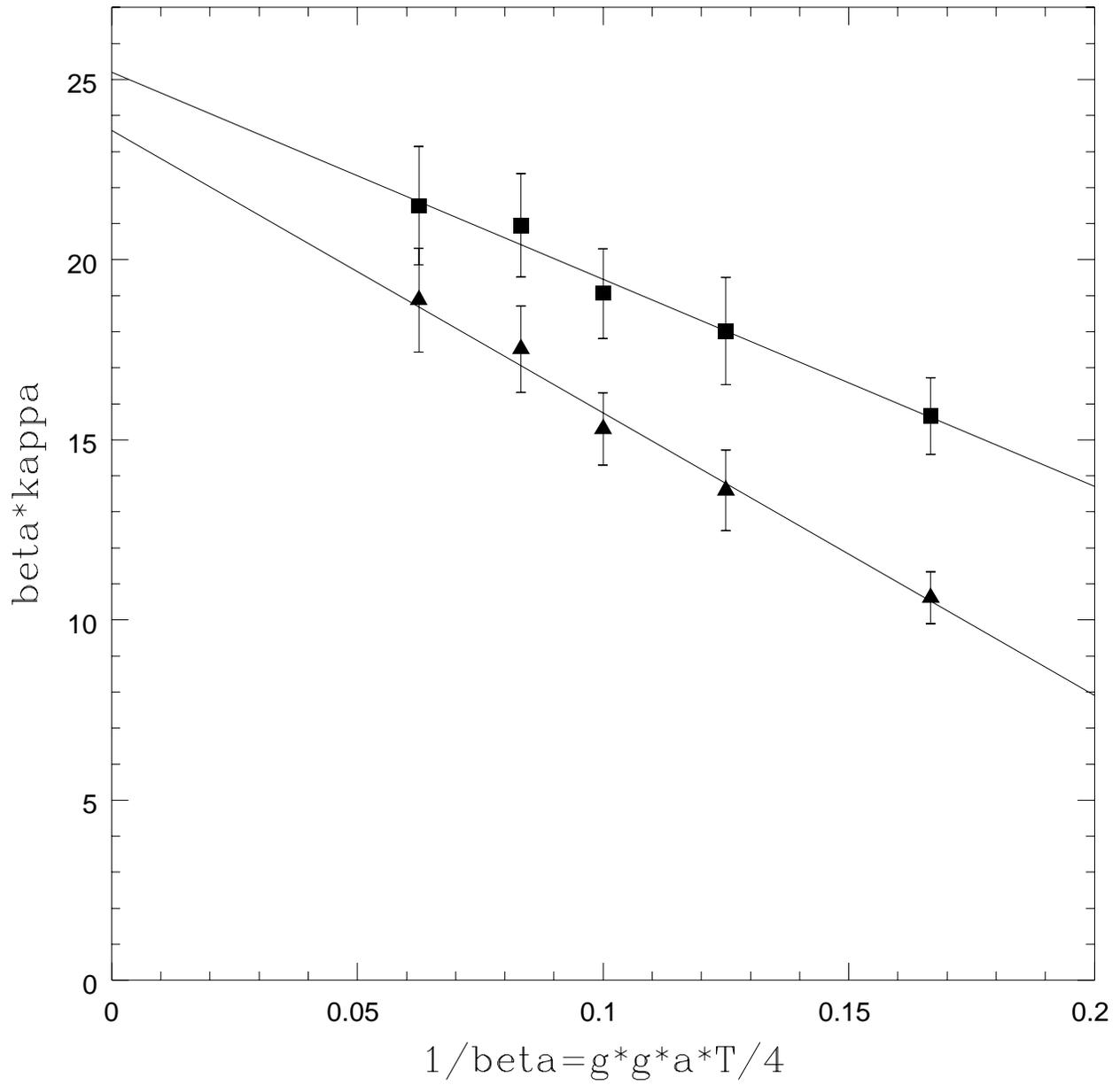}}
\caption{\label{newfig} $\bar{\kappa} \times \beta_L$ (triangles) 
and $\kappa \times \beta_L$ (squares)
plotted against $\beta_L^{-1}$.  The arguments of Arnold, Son,
and Yaffe imply a good small $\beta_L^{-1}$ (small $a$) limit.  We
display a linear fit; the extrapolation seems plausible but the fit
should not be overinterpreted.}
\end{figure}

It is possible that the argument of Arnold, Son and Yaffe is correct in
the strict $\alpha_W \rightarrow 0$ (for our purposes, $\beta_L \rightarrow
\infty$) limit, but that the lattices investigated represent 
an intermediate value where their approximations do not fully apply.
One might expect for instance that there are corrections which depend on
the ratio of $l_d$, the Debye length, and $l_n$, some 
characteristic nonperturbative length scale, leading to order
$(l_d/l_n)^2 \propto \beta_L^{-1}\sim a$ corrections to a $\kappa \propto
\beta_L^{-1}$ scaling law.  To illustrate this, we plot $\beta_L
\kappa$ against $\beta_L^{-1}$ (lattice spacing), in Figure
\ref{newfig}.  
The argument of Arnold, Son, and Yaffe is that such a plot should 
give a curve with a finite $y$ intercept; but there may be
a nonzero slope in the
approach to this point.  For illustrative purposes 
we fit our data, both with an without
thermodynamic corrections, to a straight line.  The fit of the 
data before thermodynamic corrections has a smaller slope, but this
does not mean that the thermodynamic corrections are wrong; there could
be both a non-negligible $O(a)$ thermodynamic correction and an $O(a)$
correction to the Arnold Son Yaffe scaling law, which are of opposite
sign.  The thermodynamically corrected data do not exclude the Arnold
Son Yaffe scaling law, but they require that there are 
quite large finite $(l_d/l_n)^2$ corrections.  Note that $l_d$ of the 
lattice Yang-Mills system equals the value
in the physical quantum system, $m_D^2 = 11 g^2 T^2/6$,
at $\beta_L \simeq 17$, if one uses the physical value of $g^2$; so
the strength of hard thermal loop effects in the finest lattices used
are approaching the physically interesting values.  The fit from Figure
\ref{newfig} suggests about a $20 \%$ finite $(l_d/l_n)^2$ correction from
the $\alpha_W^5$ scaling rule at this value, but the difference between
the lattice and continuum hard thermal loops means that we cannot
directly use 
the lattice value of either this correction, or of 
the coefficient for the $\alpha_W^5$ scaling law, to give the continuum
theory value.  As for verifying whether the argument of Arnold, Son, and
Yaffe is correct, our results appear consistent with their argument but
demand substantial finite $(l_d/l_n)^2$ corrections.  

It is difficult to view the data as consistent with a finite 
limit to the classical Sphaleron
rate in the $a \rightarrow 0$ limit,
ie with an $\alpha_W^4$ scaling rule, becase if the ultraviolet
physics is unimportant to the infrared dynamics (the assumption which
gives an $\alpha^4$ scaling rule) then it is hard to see why there
should be very large $O(a)$ dynamical corrections, which Figure 
\ref{resultfig} clearly requires.

\section{Conclusion}
\label{conclusion}

We have demonstrated a topological solution to
the problem of tracking Chern-Simons number in classical, real time lattice 
evolutions.  Our technique suffers from fairly substantial noise, but
this noise is white at least on long time scales and does not influence the
extracted Chern-Simons number diffusion constant per unit physical
4-volume.  

We have used our technique to study the diffusion constant for $N_{CS}$
in the broken electroweak phase and find that the rate is very substantially
suppressed with respect to the symmetric phase rate; for the Higgs 
self-coupling we used, the rate at the equilibrium temperature
is consistent with zero and is less than $\kappa = 0.005$, in agreement with
the expectations of the ``Sphaleron approximation'' estimate of the rate
but in contradiction to previous results, which were contaminated by
ultraviolet diffusion due to the non-topological definition of $N_{CS}$ used
there \cite{TangSmit}.  We have also measured the 
diffusion constant in Yang-Mills theory at a range of
lattice spacings, using an $O(a)$ improved match between the lattice 
spacing and the physical length scale, and we find a nonzero dependence on
lattice spacing, which implies that the infrared dynamics 
relevant to $N_{CS}$ diffusion depends in an important way on the physics
of hard thermal loops.  This contradicts the results of 
\cite{AmbKras,Moore1} because the definition of $N_{CS}$ used
there is screened by ultraviolet modes, and undermeasures true topology
change by a $\beta_L$ dependent amount.
The strength of the $\beta_L$ dependence in our data is weaker than that
predicted by Arnold, Son, and Yaffe
\cite{ArnoldYaffe,HuetSon}, but it is consistent with their proposal if
there are substantial finite $(l_d/l_n)^2$ corrections.

We will end by mentioning some limitations of the slave field technique
developed here, and we will comment on their possible resolution.
We will also comment briefly on the classical lattice approach in general,
in light of these results.

The most serious defect of the technique as presented here is that it
breaks down as the lattice volume becomes too large.  The reason is that
we have advocated not applying a gauge change, anywhere, whenever the slave 
field is under too much stress at any one point.  As the volume increases,
it will become ever more likely that one such site is present somewhere
in the volume; at sufficient
volume there will generically be a texture collapse event in its final
stages, somewhere in the volume, at any one time.  This will prevent gauge
changes from ever being made, or at least they will become very rare; but
as we have discussed, the gauge will then drift away from being smooth and
the winding number determination will become unstable.  In a large volume
it is necessary, then, to make gauge transformations which do not change
the gauge very near places with high slave field stress, but gauge to
slave field equals identity far away, and smoothly interpolate in between.
We have not yet implemented this idea.

The other defect of the slave field technique as presented here is the
approximation that the configuration in which the minimal energy slave field
is everywhere the identity has $N_{CS}=0$.  The white noise
we discussed is essentially due to ignoring the $N_{CS}$ of this 
configuration.  This defect could be greatly ameliorated by making a
reasonably accurate estimate of $N_{CS}$ for this configuration, every
time it is desired to write out the current value of $N_{CS}$.
While this change would not alter the diffusion constant of $N_{CS}$
measured, it might improve the statistics of the determination, since
shorter time scale diffusive motion could be seen over white noise and
included in the statistics.  Note that the technique for estimating $N_{CS}$ 
could be reasonably numerically expensive without affecting performance,
since it need only be applied every time we wish to read out $N_{CS}$, not
every time step; in particular it might be easier to find $N_{CS}$ of
a cooled version of the configuration in question.  

Finally, we should remark that since the classical, lattice rate of $N_{CS}$
diffusion is lattice spacing dependent, it is a difficult problem
to relate the lattice results to the rate in the physical, quantum 
theory.  The complications have been outlined recently in \cite{Arnoldnew}.
Probably to find the rate applicable to the quantum theory, we will
have to improve the lattice simulation in some way which properly
reproduces the hard thermal loop effects, a possibility foreseen in
\cite{BodMc}.  For instance, one could add ``particle'' degrees of 
freedom, as suggested in \cite{chaoran}.  Almost any practical, 
gauge invariant method will almost certainly need to be formulated in
terms of lattice link variables, and in this case $N_{CS}$ can be
tracked topologically using the slave field method presented here.

\vspace{0.1in}

\centerline{Acknowledgements}

\vspace{0.1in}

We wish to thank Peter Arnold, Chris Barnes, Martin Bucher, Alex
Krasnitz, Anthony Phillips, and Michael Shaposhnikov for
useful conversations or correspondence.
This research was funded in part by a PPARC (UK) rolling grant, by a 
start-up grant provided to NT by Cambridge University, and by the support
of HEFCE (UK) for the Cosmology Computing Center.
GM was supported under NSF contract NSF -- PHY96-00258.
Research was conducted in cooperation with Silicon Graphics/Cray Research 
utilizing the Origin 2000 supercomputer.

\appendix

\section{Lattice and continuum time scales}

In this appendix we estimate (but do not compute) the $O(a)$ improved 
matching condition between the lattice and continuum time scales.

First observe that there is a simple argument which almost exactly
explains the rescaling which was necessary for the spatial length scales.
The most serious difference between the lattice and continuum implementations
of the gauge theory is that the noncompact $A$ fields are replaced with
compact link matrices $U$.  The matrix $U$ is not $1 + i \tau \cdot A$
but $1 + i \tau \cdot A - A^2/2 - i \tau \cdot A A^2/6 + \ldots$, and
these higher terms generate an infinite set of extra (lattice artifact)
interaction terms.  There are also derivative corrections to all 
interaction terms which only appear in the lattice theory.  These 
properties of the lattice theory are actually inevitable in any implementation
which is gauge invariant with a local Lagrangian.  They lead to extra
diagrams in the perturbation expansion of the lattice theory and substantial
corrections in the values found for diagrams which do have continuum
analogs, leading to substantial differences between the ultraviolet 
renormalizations of the two theories (lattice and continuum), which
must be corrected for in the Lagrangian of the lattice theory.
In the case of 3 dimensional gauge theory, these corrections vanish at
least linearly in $a$; the $O(a)$ corrections arise at one loop
in lattice perturbation theory and are computed in \cite{Oapaper}.
There is a very simple mean field theory argument, due to Lepage and
Mackenzie \cite{Lepage}, which allows a quick and startlingly accurate
estimate of their size, namely the ``tadpole improvement'' technique.
The idea is to guess that the correction can be absorbed by rescaling 
each link matrix by a constant factor equal to the $-1/4$ root of 
$\langle 1/2 {\rm Tr} \Box \rangle$, the average plaquette divided by its
vacuum value.  To see how well this approximation works, compare the
$O(a)$ correction to the length scale for SU(2) theory 
with the $A_0$ field, discussed
in Section \ref{four}, with the tadpole estimate:
\begin{equation}
\beta_{L, {\rm imp}} = \beta_L - 0.701 \, , \qquad
\beta_{L, {\rm imp}} ({\rm tadpole \; estimate}) = \beta_L - 1.000 \, .
\end{equation}
Similarly, in the pure compact Abelian theory, the one loop correction
(with or without the $A_0$ field, since it does not interact) is
$\beta_{L,{\rm imp}} = \beta_L - 1/3$, which is exactly the tadpole
prediction.

We will simply assume that the dominant contribution to the rescaling of
the time scale also arises from tadpoles.  Since tadpole contributions 
appear already in the abelian theory, we can look at this (much simpler)
case to try to understand the rescaling of the time scale.

In the abelian theory, we just need to find the oscillation frequency
of an infrared mode (sufficiently infrared that we need not worry about
nonrenormalizable corrections to its dispersion relation).  The relation
\begin{equation}
\frac{d}{dt} \epsilon^i A_i(k) = \epsilon^i E_i(k)
\end{equation}
holds in the abelian theory, for transverse fields and in temporal gauge.
(Here $\epsilon^i$ is a unit vector satisfying $\epsilon^i k_i = 0$.
Henceforth we will write $\epsilon^i A_i$ simply as $A$.)
Now $\langle A^2(k) \rangle$ receives the tadpole correction and is
\begin{equation}
\langle A^2(k) \rangle = \frac{1}{k^2 \beta_{L,{\rm imp}}} \, ,
\end{equation}
while the electric field appears quadratically in the Hamiltonian and
will strictly obey equipartition,
\begin{equation}
\langle E^2(k) \rangle = \frac{1}{\beta_L} \, .
\end{equation}
Hence the frequency squared of oscillation of this mode is 
$\omega^2 = \langle E^2 \rangle / \langle A^2 \rangle = 
k^2 \beta_{L,{\rm imp}} / \beta_L$.  The time scale is shifted by
$\sqrt{\beta_L/\beta_{L,{\rm imp}}}$ with respect to the length scale,
and hence we should use $\sqrt{\beta_L \beta_{L,{\rm imp}}}$, not 
$\beta_{L,{\rm imp}}$, when we relate the physical and lattice time scales.

This argument does not apply unmodified to the case of SU(2).  For instance,
the value of $\langle A_i^2(k) \rangle$ on the lattice depends on the
renormalization of the $A$ field, which
is gauge fixing dependent, and the condition $dA/dt = E$ 
is not true in temporal gauge on the lattice.  However, these corrections
do not arise at the level of the tadpole improvement approximation, because
if they did they would be present already in the compact abelian theory;
so we anticipate that they will be subdominant compared to the quite
large $O(a)$ corrections present in $\beta_{L,{\rm imp}} - \beta_L$, and
that {\it most} $O(a)$ corrections will be absorbed by scaling between
lattice and continuum time units using $\sqrt{\beta_L \beta_{L,{\rm imp}}}$,
as in the abelian theory.  This is the approximation applied in the body
of the paper.

\begin{figure}
\centerline{\psfig{file=triang.epsi,width=\hsize}}
\caption{\label{triangulate} Triangulation into tetrahedra of two
neighboring unit cubes, with the verticies of the tetrahedra, correctly
ordered, listed below them.}
\end{figure}

\end{document}